\documentstyle[12pt,aaspp]{article}
\begin{document}
\title{The Growth Rate of Tidally Excited Waves in Accretion
Disks}
\author{Ethan T. Vishniac and Changsong Zhang}
\affil{Department of Astronomy, University of Texas, Austin TX
78712\ \ I:ethan@astro.as.utexas.edu}
\begin{abstract}
Accretion disks in close binary systems are subject to a tidally
driven parametric instability which leads to the growth of internal waves
near the outer edges of such disks (\cite{g93}). These waves
are important in understanding the torque exerted on a disk by tidal
forces and may play a role in the structure of the disk at small
radii.  Here we calculate the growth rate of this instability, including
the effects of vertical structure and fluid compressibility.
We find growth rates which are only slightly
different from Goodman's original results, except that near the
vertical resonance radius the growth rate can have an extremely broad
and strong peak when the disk is stably stratified in the vertical
direction.  Higher order modes, in the sense of increasing
number of vertical nodes, have similar growth rates.
Our results differ from a previous calculation
along these lines by Lubow et al. (1993).  The difference is
mostly due to their neglect of radial gradients in the tidally
distorted streamlines.
\end{abstract}
\section{Introduction}

In a recent paper, Goodman (1993) pointed out that a tidally
perturbed accretion disk would be subject to a parametric
instability driven by the ellipticity of the distorted
flow lines.  This instability drives the creation of a standing
wave pattern of low frequency waves with an amplitude that
rises sharply near the outer edge of the disk.  These waves
are G-modes, or buoyancy waves, whose restoring force comes
partly from vertical stratification of the fluid due to the
vertical entropy gradient and partly from the radial
stratification in the disk due to the steep radial gradient in
specific angular momentum.  Waves dominated by the former
force are usually referred to as G-modes or (in a terrestrial
context) internal waves.  Waves dominated by the latter force
are usually referred to as inertial waves.  Here we will refer
to them as internal waves.  An overview of their properties
in disks can be found in Vishniac and Diamond (1994, see also
the references contained therein).

The generation of these waves is important for at least three
reasons.  First, the waves created in this manner carry negative
angular momentum, relative to the angular momentum of the disk,
and so their generation corresponds to a torque on the disk.
This torque acts as a sink for the angular momentum transported
outward in an accretion disk and will be important in determining
its outer radius.  Second, this process will lead to a strong
inward flux of such waves in regions where the tidal forces
are weak.  The dissipation of these waves at small radii provides
a weak, but potentially significant mechanism for angular
momentum transport (\cite{vd89}).  Third, these waves can
generate dynamo activity and angular momentum transport which,
in an ionized disk, can dominate the angular momentum transport
mediated directly by the waves (\cite{vjd90}, \cite{vd92}).
This last process will be insignificant if the Balbus-Hawley
instability drives a dynamo with a growth rate comparable
to $\Omega$ (\cite{bh91}) but this claim is inconsistent with
phenomenological models of ionized accretion disks in binary
systems (see, for example \cite{s84a}, \cite{s84b}, \cite{mmh84},
\cite{hw89}, \cite{mw89}, \cite{mw89a}, \cite{c94}, \cite{ccl95}).

The physical basis of instability is closely related to a
three dimensional instability of an incompressible fluid
with elliptical flow lines (\cite{p86}, \cite{b86}, \cite{ls87}).
An observer moving with the fluid in an accretion disk will see the
tidal forces as inducing a set of time-dependent stresses
whose comoving frequency, $\bar\omega$, is $m(\Omega-\Omega_{b})$,
where $m$ is the dimensionless azimuthal wavenumber of the tidal
force, $\Omega$ is the local orbital frequency in the disk, and
$\Omega_b$ is the binary orbital frequency.  A realistic tidal
force will have components at all values of $m$, but the strongest
components will have small values of $m$.  If we consider the
$m=2$ component then we see that an internal wave with
$\bar\omega=\Omega-\Omega_b$ and $m=1$ can interact with this component
of the tidally induced stress to produce a wave traveling in the
opposite direction, with the same $m$ and $\bar\omega$.  In Goodman's
paper he showed that this process could lead to exponential
growth as each component of the standing wave amplifies the other.
Since this process is inherently local no special boundary conditions
are required.  It {\it is} necessary that the waves be confined
to the midplane of the disk, since the waves will need to maintain
coherence over a radial distance comparable to their group velocity
divided by the instability growth rate, but internal waves are
naturally confined in this manner.

The significance of this instability depends on its growth rate.
A small growth rate not only implies weaker waves, but makes the
instability vulnerable to suppression from dissipative effects,
including turbulent dissipation from any other instabilities present.
Goodman (1993) calculated the growth rate to leading order in the
tidal forces for an adiabatic fluid in a disk without
vertical structure assuming that the induced fluid motions were
incompressible.  He found that at small $r$ the growth rate
was given by
\begin{equation}
\sigma(r)\approx {15\over 4}{GM_2\over a^3\Omega(r)},
\end{equation}
where $M_2$ is the mass of the secondary (mass-losing)
star, and $a$ is the semi-major axis of the binary orbit.
This result is consistent with the growth rate found by
Ryu and Goodman (1994), who used a different approach
and who also carried out a two dimensional simulation
to look at the nonlinear saturation of the instability.
Goodman also found a singularity in the growth rate near
the vertical resonance discovered by Lubow (1981). This
resonance occurs at the radius where
\begin{equation}
\Omega(r)={\Omega_b\over 1-(\gamma+1)^{1/2}/2},
\label{eq:vertres}
\end{equation}
where $\gamma$ is the adiabatic index of the gas in the disk.
This singularity is due to pumping of the waves by vertical
motions rather than the ellipticity of the flow lines, but is a
natural part of the parametric instability in accretion disks.

More recently, Lubow, Pringle, and Kerswell (1993)
repeated the growth rate calculation including the effects
of an isothermal vertical structure.  They found a
reduced growth rate at all radii, by about an order of magnitude,
which they ascribed to the stabilizing effects of the vertical
stratification.  However, their calculation differed in other
respects as well.  In particular, they neglected vertical
motions, which eliminated the singularity, and they treated the
flow lines as locally self-similar, which eliminated radial
derivatives of the tidal distortions.  They also suggested that
the instability could be suppressed by a suitable choice of
radial boundary conditions, but further work by Ryu et al. (1995)
has shown that the instability is purely local and relatively
insensitive to radial boundaries.

In this paper we present an improved calculation of the linear
growth rate of this instability, including the effects of
vertical structure, compressibility , and variations in the local
adiabatic index
$\gamma$, while retaining tidally induced vertical motions
and radial derivatives of flow lines.  We show that a full
calculation is reasonably consistent with Goodman's original
estimate for the growth rate of the parametric instability in a realistic
disk.  However, there is one interesting difference.  Vertical
stratification greatly increases the amplitude and width of the
peak in the growth rate associated with the vertical resonance.

In \S 2 we derive the necessary formulae for calculating the
linear growth rate.  In \S 3 we present our results for disks
with different vertical structures and adiabatic indices and
compare our results with previous work.
In \S 4 we discuss the implications of this work and some
directions for future research.

\section{The Growth Rate Formalism}

The response of an accretion disk to tidal perturbations is
an example of the more general situation where one has a set
of modes of a system and one perturbs the evolution operator.
In this case the basic modes are the $m=\pm 1$ standing
internal waves of an accretion disk and the perturbation is
the change in the evolution operator for these waves due to the
tidal distortions.  If a frequency eigenstate of the system is represented
by some state vector $A_i$ then its unperturbed evolution is
given by
\begin{equation}
\partial_t A_i=L_{ij} A_j=i\bar\omega A_j,
\end{equation}
where we have used implicit summation and $\bar\omega$ is the
comoving frequency, i.e. we treat the mode evolution from the
point of view of an observer moving with the fluid.   It is
useful to define an adjoint state vector and corresponding
eigenmodes, $\tilde A_i$, using the adjoint of $L_{ij}$, i.e.
\begin{equation}
\partial_t \tilde A_i=-L_{ij}^{\dagger}\tilde A_j=i\bar\omega \tilde A_j.
\end{equation}
When $iL$ is a hermitian operator the two sets of eigenvectors are
identical.  We can define an inner product, $\langle \tilde A A\rangle$,
in the usual way as
\begin{equation}
\langle \tilde A A\rangle\equiv \int \tilde A^*_i A_i dV,
\end{equation}
where the integral is over the volume of the system.

When the system is perturbed so that $L\rightarrow L+\Delta L$
then the eigenfunctions are perturbed as well.
We can invoke the normalization constraint, i.e.
\begin{equation}
\langle \tilde A A\rangle =
\langle (\tilde A+\tilde \Delta)(A+\Delta)\rangle,
\end{equation}
to show that
\begin{equation}
\langle\tilde A+\tilde\Delta|L+\Delta L|A+\Delta\rangle=
i(\bar\omega+\delta\omega)\langle \tilde A A\rangle=i\bar\omega\langle
\tilde A A\rangle +\langle\tilde A|\Delta L|A\rangle,
\end{equation}
or
\begin{equation}
i\delta\omega={\langle\tilde A|\Delta L| A\rangle\over \langle \tilde
A A\rangle}.
\label{eq:gr1}
\end{equation}
In order to solve for the growth rate we need to find the unperturbed
eigenvectors $A$ and $\tilde A$ and calculate the matrix elements due to
tidal perturbations of the disk.

We will assume that we are dealing with standing waves whose radial
structure can be approximated by a plane wave of the form $\cos(k_r r+\phi)$,
where $\phi$ is an arbitrary phase function, and
whose azimuthal wavenumber is $\pm 1/r$.  The latter is a requirement
for an exact resonance with the $m=2$ component of the tidal force.
More generally we do not need an exact resonance to drive the instability,
but we will restrict ourselves to that case in this paper.
Waves with a low $k_\theta r \equiv m$ of order unity will show
secular radial variations on scales $\sim r/m$ (\cite{vd89}).
As long as we restrict ourselves to $k_r \gg k_\theta$, as we do here,
then we can ignore this effect.

We will restrict ourselves to adiabatic perturbations of a disk
with arbitrary vertical structure.  The evolution of radial
plane wave perturbations of the disk, with radial wavenumber
$k_r$, are described by the following equation.
\begin{equation}
{d\over dt}\pmatrix{ v_r\cr
 v_\theta\cr
 v_z\cr
 \chi\cr
 \delta}
=\pmatrix{0&2\Omega&0&-Sik_r&0\cr
-\Omega/2&0&0&0&0\cr
0&0&0&-S[(\ln(S\rho))'+\partial_z]&-z\Omega^2\cr
-{c_s^2\over S}ik_r&0&{1\over S}(z\Omega^2-c_s^2\partial_z)&0&0\cr
-ik_r&0&-(\ln(\rho)'+\partial_z)&0&0}
\pmatrix{v_r\cr
v_\theta\cr
v_z\cr
\chi\cr
\delta},
\label{eq:state1}
\end{equation}
where a prime denotes a
derivative in the $\hat z$ direction, $S\equiv P^{1/\gamma}/\rho$,
$c_s(z)$ is the sound speed, $\rho(z)$ is the density, $P(z)$ is
the pressure, $\delta$ is the fractional density perturbation,
and $\chi$ is the pressure perturbation divided by $P^{1/\gamma}$.

If we take the perturbation to have a well-defined comoving frequency,
consistent with our approximation of the perturbation as a plane
wave, we get the following second order equation in $z$ for $\chi$
\begin{equation}
\left(\chi'\ln\left[{S^2\rho\over N^2-\bar\omega^2}\right]\right)'
+(\bar\omega^2-N^2)\left({1\over c_s^2}+{k_r^2\over\Omega^2-\bar\omega^2}
\right)\chi=0,
\label{eq:struct1}
\end{equation}
where $N^2$ is the square of the Brunt-V\"ais\"al\"a frequency, defined by
\begin{equation}
N^2\equiv z\Omega (\ln S)'.
\end{equation}
This equation has an apparent singularity when $\bar\omega^2=N^2$,
but closer examination shows that there are always two linearly
independent and regular solutions passing through this point.
The only boundary condition is the requirement that $\chi$ vanish
as $|z|\rightarrow\infty$.  The other variables can be found in
terms of $\chi$ using equation (\ref{eq:state1}).  Since terms of
order $m/(rk_r)$ are negligible, they are
\begin{equation}
v_r={-k_r\bar\omega S\over\bar\omega^2-\Omega^2}\chi,
\label{eq:vr1}
\end{equation}
\begin{equation}
v_\theta={-ik_rS\over\bar\omega^2-\Omega^2}{\Omega\over2}\chi,
\end{equation}
\begin{equation}
v_z={i\bar\omega S\over\bar\omega^2-N^2}\partial_z\chi,
\end{equation}
and
\begin{equation}
\delta={\partial_z S\over\bar\omega^2-N^2}\partial_z\chi+{S\over c_s^2}
\chi.
\label{eq:d1}
\end{equation}
There will be an infinite number of solutions that satisfy this
boundary condition, just as there are an infinite number of harmonics
of the resonant mode of a box.  However, modes with many vertical
nodes, or alternatively modes with large $k_z$, will be more easily
saturated due to nonlinear processes.  Here we will consider only
a few of the lowest order modes.
In order to study the parametric instability we need to look at
the standing wave pattern formed by
the superposition of outgoing and ingoing waves.  For a given
$\bar\omega$ and $k_r$ this involves an overall phase factor of
the form
\begin{equation}
\chi\propto \cos(k_r r+\phi)\cos(\bar\omega t+\theta+\psi),
\label{eq:phase}
\end{equation}
where $\phi$ and $\psi$ are phase constants and $\theta$ is
the azimuthal coordinate.  The angle $\phi$
is irrelevant and may vary from one annular region to
another.  (The instability only requires a limited amount of
radial coherence.)  Phase factors for the other components of
the state vector can be derived from equations (\ref{eq:vr1})
through (\ref{eq:d1}).  Given a model for the vertical structure
of a disk the vertical eigenfunctions can be calculated from
these equations in a straightforward manner.

The evolution of the adjoint state vector is given by
\begin{equation}
{d\over dt}\pmatrix{\tilde v_r\cr
\tilde v_\theta\cr
\tilde v_z\cr
\tilde \chi\cr
\tilde \delta}
=\pmatrix{0&{\Omega\over 2}&0&-ik_r{c_s^2\over S}&-ik_r\cr
-2\Omega&0&0&0&0\cr
0&0&0&-{(2-\gamma)\over S}z\Omega^2-{c_s^2\over S}\partial_z&(\ln\rho)'-
\partial_z\cr
-ik_rS&0&S((\ln\rho)'-\partial_z)&0&0\cr
0&0&z\Omega^2&0&0}
\pmatrix{\tilde v_r\cr
\tilde v_\theta\cr
\tilde v_z\cr
\tilde \chi\cr
\tilde \delta},
\label{eq:state2}
\end{equation}
where the tilde variables are defined by this equation.  Note that we
have used the relation
\begin{equation}
\partial_z(c_s^2/S)=-{(\gamma-1)\over S}z\Omega^2
\end{equation}
in deriving equation (\ref{eq:state2}).  The vertical structure equation,
analogous to equation (\ref{eq:struct1}) is
\begin{equation}
\partial_z^2\left({c_s^2\over S}\tilde\chi\right)-\partial_z\left({c_s^2\over
S}\tilde\chi\right)\left(\ln(G\rho)\right)'+{c_s^2\over S}\tilde\chi
\left({-\Omega^2\over c_s^2}G+{k_r^2\Omega^2\over\bar\omega^2-\Omega^2}
\left(1-{\bar\omega^2\over\Omega^2}+z\partial_z\left(\ln\left({z\over
G\rho}\right)\right)\right)\right)=0,
\label{eq:struct2}
\end{equation}
where
\begin{equation}
G\equiv  1-{\bar\omega^2\over\Omega^2}-{z^2k_r^2\Omega^2\over \Omega^2-
\bar\omega^2}.
\end{equation}
Equation (\ref{eq:struct2}) has an apparent singularity at $G=0$.
However, as before, one can show that there exists two linearly independent
and well-behaved solutions that pass through this point.
The other adjoint variables are given by
\begin{equation}
\tilde v_r={-k_r\bar\omega\over\bar\omega^2-\Omega^2}
\left({c_s^2\tilde\chi\over S}\left(1-{\Omega^2k_r^2z^2\over
G(\bar\omega^2-\Omega^2)}\right)-{z\over G}\partial_z\left({c_s^2\over S}
\tilde\chi\right)\right),
\label{eq:vr2}
\end{equation}
\begin{equation}
\tilde v_\theta={-i2k_r\Omega\over G(\bar\omega^2-\Omega^2)}\left(
{c_s^2\over S}\tilde\chi\left(1-{\bar\omega^2\over\Omega^2}\right)-
z\partial_z\left({c_s^2\over S}\tilde\chi\right)\right),
\end{equation}
\begin{equation}
\tilde v_z={-i\bar\omega\over\Omega^2G}\left(\left({c_s^2\tilde\chi
\over S}\right){\Omega^2 k_r^2 z\over\bar\omega^2-\Omega^2}+
\partial_z\left({c_s^2\over S}\tilde\chi\right)\right),
\end{equation}
and
\begin{equation}
\tilde\delta=-{z\over G}\left(\left({c_s^2\tilde\chi\over S}\right)
{\Omega^2k_r^2 z\over\bar\omega^2-\Omega^2}+\partial_z\left(
{c_s^2\over S}\tilde\chi\right)\right).
\label{eq:d2}
\end{equation}

These equations can be solved, once again invoking $\tilde\chi\rightarrow 0$
as $|z|\rightarrow\infty$.  Alternatively, we can notice that there is
a close relationship between the state vector and the adjoint state
vector.  The fact that $\langle \tilde A A\rangle$ is a constant suggests
that we should be able to equate the state norm with some conserved
state property, e.g. the energy.  In fact, if we compare
$\tilde v_r$ and $(1/2)\rho v_r$ then we can show that equations
(\ref{eq:struct1}), (\ref{eq:vr1}), (\ref{eq:struct2}), and
(\ref{eq:vr2}) imply that the two variables satisfy the same
second order differential equation in $z$.  The boundary condition
at $|z|\rightarrow\infty$ does not distinguish between solutions
that differ only by a complex factor, but the phase and amplitude
of the adjoint state vector is undefined in any case.  We conclude
that we can take $\tilde v_r=(1/2)\rho v_r$.  With this choice,
we can use equations (\ref{eq:vr1}) through (\ref{eq:d1}) and
(\ref{eq:vr2}) through (\ref{eq:d2}) to show that
\begin{equation}
\pmatrix{\tilde v_r\cr \tilde v_\theta\cr \tilde v_z\cr
\tilde \chi\cr \tilde \delta}=\pmatrix{{1\over2}\rho v_r\cr
2\rho v_\theta\cr {1\over2}\rho v_z\cr {1\over2} \rho\left(
{S^2\over c_s^2}\chi-{S\over c_s^2}z\Omega^2\Delta z\right)\cr
{1\over 2}\rho z\Omega^2\Delta z},
\label{eq:admean}
\end{equation}
where
\begin{equation}
\Delta z\equiv {v_z\over i\bar\omega}.
\end{equation}
The meaning of this choice is clearer if we note that equation
(\ref{eq:admean}) implies that
\begin{equation}
\langle\tilde A A\rangle = {1\over 2}\rho\left(|v_r|^2+4|v_\theta|^2
+|v_z|^2\right) +{|\delta P|^2\over 2\rho c_s^2}-{1\over 2}
\rho z\Omega^2\Delta z^*\left({\delta P\over\gamma P}-\delta\right).
\end{equation}
{}Furthermore
\begin{eqnarray}
-{1\over 2}\rho z\Omega^2 \Delta z^*\left({\delta P\over\gamma P}-\delta\right)
=&-{1\over 2}\rho z\Omega^2 \Delta z^*\left({\delta S\over S}\right)\cr
=&-{1\over 2}\rho z\Omega^2 \Delta z^*\left(-\Delta z\partial_z\ln S\right)
\cr
=&{1\over 2}\rho N^2|\Delta z|^2.
\end{eqnarray}
Therefore,
\begin{equation}
\langle\tilde A A\rangle = {1\over 2}\rho\left(|v_r|^2+4|v_\theta|^2
+|v_z|^2\right) +{|\delta P|^2\over 2\rho c_s^2}+
{1\over 2}\rho N^2|\Delta z|^2,
\end{equation}
which is to say that using this convention the norm of the state is
the locally measured energy density.  Goodman (1993) used the local
energy density in his calculation of the perturbation growth rate,
but, consistent with his neglect of compressibility and
vertical structure, left out the last two terms.

It remains to find the operator $\Delta L$ and calculate the
matrix elements $\langle\tilde A|\Delta L| A\rangle$.
We will start with the lagrangian displacements produced
by the tidal forces.  We follow Goodman in using first-order
epicyclic theory to calculate the radial and angular displacements.
The radial and azimuthal displacements are
\begin{equation}
\xi_r={1\over 4\bar\omega^2-\Omega^2}\left({d\Phi_2\over dr}
+{2\Omega\over r\bar\omega}\Phi_2\right) \cos(2\bar\omega t+2\theta),
\end{equation}
and
\begin{equation}
\xi_\theta={-1\over2(4\bar\omega^2-\Omega^2)}\left(
{2\Omega\over r\bar\omega}{d\Phi_2\over dr}+{4\bar\omega^2+3\Omega^2\over
r^2\bar\omega^2}\Phi_2\right) \sin(2\bar\omega t+2\theta),
\end{equation}
where $\bar\omega=\Omega-\Omega_b$, $\theta$ is measured relative
to the axis running between the two stars, and $\Phi_2$ is the
azimuthal $m=2$ Fourier component of the tidal potential.  It
is given by (\cite{bc61})
\begin{equation}
\Phi_2=-q{GM_1\over a}b_2^{(1/2)}(r/a),
\end{equation}
where $q$ is the mass ratio for the system, $M_1$ is the primary
stellar mass, and $b_2^{1/2}(r/a)$ is a Laplace coefficient.
The vertical displacement follows from examining the response
of the disk to long radial wavelength perturbations
to the local vertical gravity.  This gives
\begin{equation}
\xi_z={z\cos(2\bar\omega+2\theta)\over 4\bar\omega^2-(\gamma+1)\Omega^2}\left
((\gamma-1)\Omega^2\left({1\over r}\partial_r (r\xi_r)+
{1\over r}\partial_\theta\xi_\theta\right)-{1\over r}{d\over dr}\left(
{d\Phi_2\over dr}\right)+{4\over r^2}\Phi_2-{3\xi_r\over r}\Omega^2
\right).
\label{eq:xiz}
\end{equation}
This differs from the expression in Goodman (1993) in two respects.
{}First, the sign of the $\Phi_2$ terms are reversed.  Second,
there is an additional term proportional to $\xi_r$ which comes
from the change in vertical gravity induced by radial motions.
Writing these displacements as a set of coefficients of the angular
functions $\cos(2\bar\omega t+2\theta)$ and $\sin(2\bar\omega t+2\theta)$
we find
the tidally induced eulerian velocity perturbations are
\begin{equation}
\Delta v_r=(\Omega \xi_\theta-2\bar\omega\xi_r)\sin(2\bar\omega t+2\theta)
\end{equation}
\begin{equation}
\Delta v_\theta=\left({\Omega\over2}\xi_r+2\bar\omega\xi_\theta\right)
\cos(2\bar\omega t+2\theta)
\end{equation}
and
\begin{equation}
\Delta v_z=-2\bar\omega\xi_z\sin(2\bar\omega t+2\theta).
\end{equation}
In addition, we need to know the tidally induced perturbations
to the pressure, density, and entropy.  These are
\begin{eqnarray}
{\Delta S\over S}=&-\xi\partial_z \ln S\cr
=&{-\xi_z N^2\over z\Omega^2}\cr
=&-{\partial_z\xi_z\over\Omega^2}N^2\cos(2\bar\omega t+2\theta),
\label{eq:ts}
\end{eqnarray}
\begin{eqnarray}
{\Delta\rho\over\rho}=&-\vec\nabla\cdot\vec\xi-\xi_z\partial_z\ln\rho\cr
=&-\left({1\over r}\partial_r(r\xi_r)+{2\over r}\xi_\theta+
\partial_z\xi_z\left(1+z\partial_z\ln\rho\right)\right)\cos(2\bar\omega t+
2\theta)
\label{eq:td}
\end{eqnarray}
and
\begin{equation} {\Delta P\over P}=
\gamma\left({\Delta S\over S}+{\Delta\rho\over\rho}\right)
\label{eq:tp}
\end{equation}
Here we have used the assumption that the tidal perturbations
are adiabatic.
This assumption may not always be valid, but since we have already
employed it in deriving the dispersion relation for the waves, we
have not further compromised our results by using it here.
In any case, this assumption should be reasonable as long as the
cooling time for the disk is much greater than the orbital time.
The radial and vertical derivatives of these expressions can be
calculated in a straightforward manner.

It is convenient to rewrite the components of the state vectors as
coefficients of phase factors, following equation (\ref{eq:phase}).
{}For both the state vector and its adjoint the phase factors have
the form
\begin{equation}
\pmatrix{\sin(k_rr+\phi)\sin(\bar\omega t +\theta+\psi)\cr
-\sin(k_rr+\phi)\cos(\bar\omega t+\theta+\psi)\cr
-\cos(k_rr+\phi)\sin(\bar\omega t+\theta+\psi)\cr
\cos(k_rr+\phi)\cos(\bar\omega t+\theta+\psi)\cr
\cos(k_rr+\phi)\cos(\bar\omega t+\theta+\psi)}.
\end{equation}
The definitions of the state vector and adjoint state vector
components are then modified by dividing through by $i$ for
every factor of $\bar\omega$ or $k_r$.  Note that when we are
done constructing our standing wave patterns there is no
distinction left between $\tilde A$ and $\tilde A^*$.
{}Finally, we note that matrix elements which involve products
of the form $\cos(k_r r)\sin(k_r r)$ will not vanish exactly.
Instead,
\begin{equation}
\langle \sin(k_r r) F(r) \cos(k_r r)\rangle =-{1\over4}\langle
k_r^{-1}\partial_r F(r)\rangle.
\end{equation}
When the matrix element involves a factor of $k_r$ such terms
will contribute to the total growth rate at the same order as
other terms.  This point is particularly significant since
the radial gradients of the tidal perturbations can be much
larger than the tidal perturbations divided by $r$.  We note
that in performing the necessary integration by parts we have
neglected secular radial variations in the eigenmodes,
which will arise from radial variations in density and pressure.

We are now ready to write down expressions for the matrix elements
which enter into equation (\ref{eq:gr1}).  The tidally perturbed
radial acceleration equation is
\begin{eqnarray}
\partial_t v_r=2\Omega v_\theta-&{1\over\rho}\partial_r P
-\left(\Delta v_r\partial_r+{\Delta v_\theta\over r}\partial_\theta
+\Delta v_z\partial_z+\partial_r\Delta v_r\right) v_r
+\left({2\Delta v_\theta\over r}-{1\over r}\partial_\theta\Delta v_r\right)
v_\theta\cr
&-v_z\partial_z\Delta v_r+{\Delta\rho\over\rho}S\partial_r\chi+
{1\over\rho}\partial_r\Delta P.
\end{eqnarray}
Taking into account the radial phase factors given in equation
(\ref{eq:phase}) we can use this to write the top row of $\Delta L$
as
\begin{equation}
\Delta L_{ri}=\left(-{1\over2}\partial_r\Delta v_r-\Delta v_z\partial_z
-{\Delta v_\theta\over r}\partial_\theta, {2\Delta v_\theta\over r}
-{1\over r}\partial_\theta \Delta v_r, 0,{\Delta\rho\over\rho}S\partial_r,0
\right).
\end{equation}
This gives a contribution to the growth rate, following equation (\ref{eq:gr1})
of
\begin{eqnarray}
\langle \tilde v_r|\Delta L| A\rangle =&{\sin(2\psi)\over 8}\int_{-\infty}
^{\infty} \tilde v_r\biggl(v_r\left(-{\partial_r\Delta v_r\over 2}
-{\Delta v_\theta\over r}\right)-\Delta v_z\partial_z v_r-\left(
{2\Delta v_\theta\over r}-{2\Delta v_r\over r}\right)v_\theta\cr
&-\chi k_r{\Delta\rho\over\rho}S\biggr)dz\cr
=&{\sin(2\psi)\over8}\int_{-\infty}^{\infty}\tilde v_r\biggl(
v_r\left(\bar\omega\partial_r\xi_r-{\Omega\over2}\partial_r\xi_\theta
+\left({3\over4}\Omega-2\bar\omega\right){\xi_\theta\over r} -2\Omega
{\xi_r\over r}\right)+\partial_z v_r2\bar\omega\xi_z\cr
&-{2v_\theta\over r}\left(\xi_r\left({\Omega\over 2}+2\bar\omega
\right)+\xi_\theta\left(2\bar\omega-\Omega\right)\right)\cr
&+\chi k_r S\left({\xi_r\over r}+\partial_r\xi_r+{2\over r}\xi_\theta
+\partial_z\xi_z\left(1+z\partial_z\ln\rho\right)\right)\biggr)dz.
\label{eq:1row}
\end{eqnarray}

Similarly, the tidally perturbed azimuthal acceleration equation is
\begin{eqnarray}
\partial_tv_\theta=&-{\Omega\over2}v_r-\left(\partial_r\Delta v_\theta
+{\Delta v_\theta\over r}\right)v_r-\left(\Delta v_r\partial_r
+{\Delta v_\theta\over r}\partial_\theta+\Delta v_z\partial_z
+{\Delta v_r\over r}+{\partial_\theta\Delta v_\theta\over r}\right)v_\theta
\cr
&-v_z\partial_z\Delta v_\theta,
\end{eqnarray}
which implies a contribution to the growth rate of
\begin{eqnarray}
\langle \tilde v_\theta|\Delta L| A\rangle =&{\sin(2\psi)\over 8}\int_{-\infty}
^{\infty} \tilde v_\theta\biggl(v_r\left(\partial_r\Delta v_\theta+{\Delta
v_\theta\over r}\right)+v_\theta\left(-{1\over2}\partial_r\Delta v_r
+{\Delta v_r\over r}-{\Delta v_\theta\over r}\right)\cr
&+\Delta v_z\partial_z
v_\theta\biggr)dz\cr
=&{\sin(2\psi)\over 8}\int_{-\infty}^{\infty}\tilde v_\theta\biggl(
v_r\left({\Omega\over2}\partial_r\xi_r+2\bar\omega\partial_r\xi_\theta
-{\Omega\over 4}{\xi_r\over r}+(2\bar\omega-3\Omega){\xi_\theta\over r}\right)
\cr
&+v_\theta\left(-{\Omega\over2}\partial_r\xi_\theta+\bar\omega\partial_r\xi_r
+\left({7\over4}\Omega-2\bar\omega\right){\xi_\theta\over r}
-2(\Omega+\bar\omega){\xi_r\over r}\right)\cr
& -2\bar\omega\xi_z\partial_z v_\theta\biggr)dz.
\label{eq:2row}
\end{eqnarray}

The tidally perturbed vertical acceleration equation is
\begin{eqnarray}
\partial_t v_z=&-S\left(\partial_z\ln\left({S\over\rho}\right)
+\partial_z\right)\chi-z\Omega^2\delta\cr
&-v_r\partial_r\Delta v_z-{1\over r}\partial_\theta\Delta v_z
-v_z\partial_z\Delta v_z -\left(\Delta \vec v\cdot\vec\nabla\right)
v_z\cr
&+\Delta\rho S\left(-{z\Omega^2\over c_s^2}+\partial_z\right)\chi
+{\delta\over\rho}\partial_z\Delta P+2\delta{\Delta\rho\over\rho}z\Omega^2,
\end{eqnarray}
which implies a contribution to the growth rate of
\begin{eqnarray}
\langle \tilde v_z|\Delta L| A\rangle =&{\sin(2\psi)\over 8}\int_{-\infty}
^{\infty} \tilde v_z\biggl(v_z\left(-\partial_z\Delta v_z+{1\over2}
\partial_r\Delta v_r-{\Delta v_\theta\over r}\right)-\Delta v_z\partial_z
v_z\cr
&-{\Delta\rho\over\rho}S\left(-{z\Omega^2\over c_s^2}+\partial_z\right)\chi
-\delta\left({1\over\rho}\partial_z\Delta P+2{\Delta\rho\over\rho}z\Omega^2
\right)\biggr)dz\cr
&={\sin(2\psi)\over8}\int_{-\infty}^\infty\tilde v_z\biggl(v_z\left(
{\Omega\over2}\partial_r\xi_\theta-\bar\omega\partial_r\xi_r+\Omega{\xi_r
\over r}-\left({3\over4}\Omega+2\bar\omega\right){\xi_\theta\over r}
+2\bar\omega\partial_z\xi_z\right)\cr
&+2\bar\omega\xi_z\partial_zv_z-{\Delta\rho\over\rho}S\left(
-{z\Omega^2\over c_s^2}+\partial_z\right)\chi\cr
&+z\Omega^2\left((\gamma-2){\Delta\rho\over\rho}+\gamma{\Delta S\over S}-
\partial_z\xi_z\left(2+z\partial_z\ln\rho+{\gamma z^2\Omega^2\over c_s^2}
\right)\right)\delta\biggr)dz.
\label{eq:3row}
\end{eqnarray}

The tidally perturbed pressure evolution equation is
\begin{eqnarray}
\partial_t\chi=&-{c_s^2\over S}\partial_r v_r+{1\over S}(z\Omega^2v_z-
c_s^2\partial_zv_z)\cr
&+P^{-1/\gamma}\biggl((-\vec v\cdot\vec\nabla)\Delta P-\gamma\Delta P
\vec\nabla\cdot\vec v-\Delta\vec v\cdot\vec\nabla(P^{1/\gamma}\chi)
-\gamma\vec\nabla\cdot(\Delta\vec v)P^{1/\gamma}\chi\biggr),
\end{eqnarray}
which implies a contribution to the growth rate of
\begin{eqnarray}
\langle \tilde\chi|\Delta L| A\rangle =&{\sin(2\psi)\over 8}\int_{-\infty}
^{\infty} \tilde\chi\biggl(-{c_s^2\over S}{\Delta P\over P}k_rv_r+
{c_s^2\over S}\left({\partial_z\Delta P\over\gamma P}+{\Delta P\over P}
\partial_z\right)v_z\cr
&\left(\left(\gamma-{1\over2}\right)\left({\partial_r(r\Delta v_r)\over r}
-{2\Delta v_\theta\over r}\right)+\Delta v_z\left(-{z\Omega^2\over
c_s^2}+\partial_z\right)+\gamma\partial_z(\Delta v_z)\right)\chi\biggr)dz.
\cr
=&{\sin(2\psi)\over 8}\int_{-\infty}^{\infty} \tilde\chi\biggl(
-{c_s^2\over S}k_r{\Delta P\over P}v_r
-{z\Omega^2\over S}\left({\Delta P\over P}-\partial_z\xi_z
\left(2+z\partial_z\ln\rho+{z^2\gamma\over c_s^2}\right)\right)v_z\cr
&+{c_s^2\over S}{\Delta P\over P}{dv_z\over dz}+
\chi\biggl(2\bar\omega\left({z^2\Omega^2\over c_s^2}-\gamma\right)
\partial_z\xi_z +\cr
&\left(\gamma-{1\over2}\right)\left(2\Omega_b{\xi_r\over r}
-\xi_\theta\left({\Omega\over2}+4\bar\omega\right)+\Omega\partial_r\xi_\theta
-2\bar\omega\partial_r\xi_r\right)\biggr)
-2\bar\omega\xi_z{d\chi\over dz}\biggr)dz
\label{eq:4row}
\end{eqnarray}

{}Finally, the tidally perturbed density evolution equation is
\begin{eqnarray}
\partial_t\delta=&-\partial_rv_r-v_z\partial_z\ln(\rho)-\partial_z v_z
-{\Delta\rho\over\rho}\left(\partial_rv_r+{v_r\over r}\right)-\partial_r
\left({\Delta\rho\over\rho}\right)v_r\cr
&-{\Delta\rho\over\rho}{1\over r}
\partial_\theta v_\theta-{v_\theta\over r}\partial_r\left({\Delta\rho
\over\rho}\right)
-{\Delta\rho\over\rho}\partial_zv_z-{v_z\over\rho}\partial_z\Delta\rho
-\delta(\vec\nabla\cdot\Delta\vec v+\Delta v_z\partial_z\ln\rho).
\end{eqnarray}
which implies a contribution to the growth rate of
\begin{eqnarray}
\langle \tilde\delta|\Delta L| A\rangle =&{\sin(2\psi)\over 8}\int_{-\infty}
^{\infty} \tilde\delta\biggl(-{\Delta\rho\over\rho}k_rv_r+
v_z{\partial_z\Delta\rho\over\rho}+{\Delta\rho\over\rho}\partial_zv_z\cr
&+\left({1\over2r}\partial_r(r\Delta v_r)-{\Delta v_\theta\over r}
+\Delta v_z\partial_z\ln\rho\right)\delta+\Delta v_z\partial_z\delta
\biggr)dz\cr
=&{\sin(2\psi)\over 8}\int_{-\infty}^{\infty} \tilde\delta\biggl(-
{\Delta\rho\over\rho}k_rv_r+v_z\left(-\partial_z\xi_z\left(
\partial_z\ln\rho-z(\partial_z\ln\rho)^2+z\partial_z^2\ln\rho\right)
+{\Delta\rho\over\rho}\partial_z\ln\rho\right)\cr
&+{\Delta\rho\over\rho}{dv_z\over dz}+\delta\left(\left(-{1\over4}\Omega
+2\bar\omega\right){\xi_\theta\over r}+(\Omega-\bar\omega){\xi_r\over r}
+{\Omega\over2}\partial_r\xi_\theta-\bar\omega\partial_r\xi_r-2\bar\omega
\xi_z\partial_z\ln\rho\right)\cr
&-2\bar\omega\xi_z{d\rho\over dz}\biggr)dz
\label{eq:5row}
\end{eqnarray}

In order to calculate the growth rate for a given vertical structure,
we need to solve for the vertical eigenfunctions using equations
(\ref{eq:struct1}) through (\ref{eq:d1}), and the adjoint eigenfunctions
using equations (\ref{eq:struct2}) through (\ref{eq:d2}) or, more
economically using equation (\ref{eq:admean}), and substitute these
results into equations (\ref{eq:1row}), (\ref{eq:2row}), (\ref{eq:3row}),
(\ref{eq:4row}), and (\ref{eq:5row}) using equations (\ref{eq:ts}),
(\ref{eq:td}) and (\ref{eq:tp}).  Finally, we need to add up
the matrix components and divide by the state norm given by
\begin{equation}
\langle\tilde AA\rangle={1\over4}\int_{-\infty}^{\infty}
\left(\tilde v_r v_r+\tilde v_\theta v_\theta
+\tilde v_z v_z+\tilde\chi\chi+\tilde\delta\delta\right)dz.
\end{equation}

\section{Results}

We are now in a position to calculate the growth rates for the
parametric tidal instability for a range of input parameters.
Although our formalism allows us to take any vertical structure
we will confine ourselves to presenting results only for adiabatic
and isothermal vertical structures.  These two cases should
give us a sense of the range of possible results.
We start with adiabatic disks.  In this case the vertical
structure is given by
\begin{equation}
P=P_0\left(1-\left({z\over H}\right)^2\right)^{{\gamma_s\over\gamma_s-1}},
\end{equation}
\begin{equation}
\rho=\rho_0\left(1-\left({z\over H}\right)^2\right)^{{1\over\gamma_s-1}},
\end{equation}
with
\begin{equation}
H=\left({2P_0\over\rho_0\Omega^2}\right)^{1/2},
\end{equation}
\begin{equation}
c_s^2=\gamma{P_0\over\rho_0}\left(1-\left({z\over H}\right)^2\right),
\end{equation}
and
\begin{equation}
N^2=0.
\end{equation}
The adiabatic index for the vertical structure is not necessarily
the same index that appears in the perturbation equations. The
latter describes the response of fluid elements on short time scales
to changes in the ambient pressure, whereas the former is an attempt to
model the vertical structure.  In the preceding equations we have
distinguished the two by using $\gamma_s$ when we mean the vertical
structure parameter.  This class of models has the peculiar property
that the boundary occurs at $|z|=H$ rather than $|z|=\infty$.
Consequently we have to modify our boundary conditions and take the
solution which is least divergent near $|z|=H$ i.e. $\chi$ goes
to a constant value near the vertical disk boundaries.  An unfortunate
consequence is that the fluid velocities are also constant, while
the sound speed goes to zero.  This is manifestly unphysical, but
not too surprising.  One expects the adiabatic approximation to
break down far from the disk midplane.  Fortunately the results of
our calculation are weighted by the mode energy, and therefore by
the local density, which drops to zero faster than the sound speed.
It seems safe to assume that adding on an isothermal atmosphere
will have little impact on our results as long as the column density
in the atmosphere remains a small fraction of the total
column density.

In figure 1 we show the growth rate for the parametric instability
in units of $q\Omega_b$ for an adiabatic disk with $\gamma_s=\gamma=5/3$.
as a function of $r/a$, where $a$ is the semi-major axis of the binary
orbit.  In this figure we plot the results for the three lowest order
modes of internal waves.  The primary mode has one zero in $\chi$ and
none in $v_z$, i.e. the disk material bobs up and down as the wave
passes by.  In fact, in the limit where $\gamma_s=\gamma=1$, $v_z$
is not a function of $z$ at all.
(However, in this case $v_r$ does have a vertical mode.
The material above and below the midplane are moving in opposite directions
radially during each cycle.)
We see that the second and third
harmonics are almost indistinguishable in this plot, and the primary
mode shows large differences only for radii close to, or greater than,
the vertical resonance radius.  The fundamental mode shows a stronger
response to the vertical resonance and a significantly larger growth rate
for larger radii.  At small radial distances our results
converge to Goodman's asymptotic formula for the higher order modes.  The
primary mode is weaker by about 23\%.  The vertical resonance peak is
basically similar to the peak in his plot, but the zero in the growth
rate occurs at a radius just outside the resonant radius, whereas
in his calculation it was on the opposite side of the peak.  The source
of this discrepancy is not clear.  It is not due to our disagreement
concerning the sign of the $\Phi_2$ terms in equation (\ref{eq:xiz}).

In figure 2 we show the growth rate in units of $q\Omega_b$ for
the fundamental mode and for $\gamma=\gamma_s=5/3, 1.2$ and $1.0$.
This last case is equivalent to taking an isothermal vertical
structure and assuming isothermal perturbations.
The two most significant effects of changing the adiabatic index are
that the vertical resonance is pushed outwards and the
peak around it is broadened.

A more realistic, but still idealized, model is to assume an
isothermal vertical structure for the disk.  In this case we
have $P\propto\rho$, $S\propto\rho^{1-1/\gamma}$,
\begin{equation}
\rho=\rho_0\exp\left({-\rho z^2\Omega^2\over 2P}\right),
\end{equation}
\begin{equation}
c_s^2=\gamma{P\over\rho},
\end{equation}
and
\begin{equation}
N^2={\rho z^2\Omega^4\over P}\left(1-{1\over\gamma}\right).
\end{equation}

In figure 3 we show the growth rate in units of $q\Omega_b$ for
the first three internal wave modes for an isothermal disk
with $\gamma=5/3$.  We see that the response to the vertical
resonance is much larger in this case, by a factor of about
$5/2$ for the fundamental mode.  This carries over into a smaller
response well beyond the radius of the vertical resonance. At
much smaller radii the growth rates are similar
to the adiabatic case, the only difference being that the primary
mode is reduced from the higher order modes even more, by about 36\%.

In figure 4 we show the growth rate in units of $q\Omega_b$ for
the fundamental mode in an isothermal disk with $\gamma=5/3, 1.2$
and $1.0$.  As before, lowering $\gamma$ broadens the vertical
resonance peak and moves it outward.

The modest differences between this calculation and Goodman (1993)
for the adiabatic disk with $\gamma=5/3$ are
explainable in terms of the additional effects included
in this paper.  On the other hand, our results are very different
from Lubow et al. (1993).  We have used a different calculational
approach, and included the tidally induced vertical motions and
pressure and density fluctuations.  However, while these effects
are quantitatively important, leaving them out does not produce
a dramatic reduction in our growth rates, except near the vertical
resonance.  The most important difference seems to be our inclusion
of radial derivatives in the tidal stresses.  We can make a crude
comparison between our two calculations by setting
$\Delta\rho=\Delta P=\xi_z=0$, and taking $\partial_r\xi_r=\xi_r/r$ and
$\partial_r\xi_\theta=\xi_\theta/r$.  In figure 5 we show the
growth rates for the fundamental mode in an isothermal disk with
$\gamma=5/3$.  The top curve shows the results of our complete calculation,
also seen in figure 3.  The next curve shows the results of removing
tidally induced vertical motions, and density and pressure fluctuations.
The bottom curve shows the results adding to this the assumption
that the streamlines are locally self-similar,
i.e. the radial derivatives of $\vec \xi$ are found by
setting $\xi_r$ and $\xi_\theta$ proportional to $r$
in the interaction matrix.  We note that at small $r$ the bottom
curve is more than a factor of 5 below the top one.
This is a dramatic reduction, although this result still lies
about 30\% above the the reduction obtained by Lubow et al..
The remaining discrepancy is apparently due to the difficulty
in entering precisely equivalent assumptions into our program,
due to our somewhat different methods for evaluating the interaction
matrix.  We conclude that the bulk of the difference between
the results of Goodman (1993) and Lubow et al. (1993) stems from the
latter's neglect of radial derivatives of the streamline perturbations,
not from their inclusion of vertical structure.

\section{Conclusions}

We have calculated the linear growth rate of a tidally induced
parametric instability in accretion disks.  Our results take
into account only leading order terms in the tidal distortion
of an accretion disk, but include compressibility, vertical
structure, vertical motions, and radial derivatives of the
tidal distortions.  Our basic conclusion is that the original
results of Goodman (1993), derived under a simpler set of
assumptions, are basically correct.  For a disk with
adiabatic vertical structure we see only small deviations
except near the vertical resonance, where we find the zero
in the growth rate lies just outside the radius of the
vertical resonance, rather than just inside it.  This difference
is unlikely to have any significant dynamical effects.
{}For a disk with isothermal vertical structure we find a
general enhancement of the growth rate near the vertical
resonance, by a factor of about $2.5$.  This implies that
the net torque on an accretion disk may depend significantly
on its vertical structure.

One lingering question regarding these results is the
meaning of the infinite growth rate for the instability
in the neighborhood of the vertical resonance.  This
singularity is
a consequence of the divergence in $\xi_z$ at that radius.
In reality there are two effects which limit this divergence.
{}First, strong vertical motions within a narrow annulus will
lead to the emission of sound waves (\cite{l81}).
{}First, equation (\ref{eq:vertres}) was derived neglecting
radial gradients.  If we define an effective radial wavenumber
for the growth rate by
\begin{equation}
K_r\equiv \partial_r \ln\sigma(r),
\end{equation}
then the tidal pumping of the vertical resonance will saturate
due to the emission of sound waves from the resonant region when
\begin{equation}
K_rV_{group}={c_s^2K_r^2\over2\bar\omega}\approx  \sigma(r).
\end{equation}
Near the vertical resonance the growth rate is approximately
\begin{equation}
\sigma(r)\approx A
\left({r\over\Delta r}\right)q{\Omega_b^2\over\Omega},
\end{equation}
where $\Delta r$ is the distance to the resonant radius and
$A$ is a function of $\gamma$ and the vertical disk structure.
We see that the half-width of the annulus within which damping
by acoustic emission is important is
\begin{equation}
\Delta r\approx{\Omega\over2(\Omega-\Omega_b)} {c_s^2\over qA(\gamma)
\Omega_b^2 r}\sim {H^2\over qrA},
\label{eq:delr1}
\end{equation}
where $H\sim c_s/\Omega$ is the disk thickness and we have ignored
the distinction between $\Omega_b$ and $\Omega$ in the last
(highly approximate) expression.
Within this annulus the effective growth rate is
\begin{equation}
\sigma_{lim}\approx {2 r^2(\Omega-\Omega_b)\over c_s^2}\left(
Aq{\Omega_b^2\over\Omega}\right)^2\sim (qA)^2 \Omega_b\left({r\over H}
\right)^2.
\end{equation}
We see that as $(H/r)\rightarrow 0$ acoustic emission becomes
ineffective at removing the effects of this singularity.

The second damping mechanism is due to nonlinear broadening of the
vertical resonance.  The resonance width will increase as
the advective terms in the acceleration equation become important.
In physical terms, the strong vertical oscillations of the disk
will excite higher order harmonics, which will interfere with the
resonant excitation of the fundamental mode.
This is a purely one dimensional effect and does not depend
on radial gradients in the tidal forcing.
We can approximate its effects by inserting a term of the form
$C_04\bar\omega^2(\partial_z\xi_z)^2$ into the denominator of
equation (\ref{eq:xiz}) where $C_0$ is a constant of order unity.
This implies a limit on $\partial_z\xi_z$ of order
\begin{equation}
(\partial_z\xi_z)_{lim}\sim \left(q{\Omega_b^2\over\Omega^2}\right)^{1/3}.
\end{equation}
Since the denominator of the leading factor in
equation (\ref{eq:xiz}) goes as $-12\Omega_b\bar\omega(\Delta r/r)$
this implies that this damping mechanism will be effective within
an annular width of approximately
\begin{equation}
\Delta r\approx r{C_0\bar\omega\over 12\Omega_b}(\partial_z\xi_z)_{lim}^2
\sim r{C_0\bar\omega\over 12\Omega_b}
\left(q{\Omega_b^2\over\Omega^2}\right)^{2/3}.
\label{eq:delr2}
\end{equation}
The instability growth rate within this annulus will be limited to
\begin{equation}
\sigma_{lim}\sim A\left(q{\Omega_b^2\over\Omega^2}\right)^{5/3}
{12\Omega_b\Omega\over C_0(\Omega-\Omega_b)}.
\end{equation}

Comparing equations (\ref{eq:delr1}) and (\ref{eq:delr2}) we note that
acoustic emission will tend to be the most important damping process
when the disk is not too
thin, or when the tidal instability is driven only weakly.  There is
no general rule as to which will be more important in accretion disks
in compact binary systems, and it is possible that for some systems
acoustic emission may dominate during outbursts while nonlinear broadening
is important during quiescence.  In either case the ratio
of the torque exerted within the annulus dominated by the vertical
resonance to the torque exerted on the rest of the disk at a similar
radius can be estimated by comparing the cube of the growth rate
weighted by the annulus width (\cite{g93}).  We have
\begin{equation}
{\sigma_{lim}^3\Delta r\over\sigma(r)^3 r}\sim
\left({4A\over15}\right)^3\left({r\over\Delta r}
\right)^2.
\end{equation}
{}For $\gamma=5/3$ $A\approx4.0$ for an isothermal disk and
$1.7$ for an adiabatic
disk.  This suggests that the vertical resonance can dominate the torque
on a disk and that accretion disks will have a tendency to have an
outer radius that lies within this annulus. This tendency should
be particularly noticeable when the disk
structure is isothermal and/or $\gamma$ is relatively small (so that
$\Omega_b/\Omega$ is relatively large).  However, apart from the
vertical resonance $\sigma(r)$ rises sharply with $r$ and the
torque per area, which goes as $H^2\sigma(r)^3$ rises even more
sharply.  This makes it possible for the dominant contribution
to the torque on a disk to come from the outer edge of the disk
provided that this lies a significantly greater radius than the
vertical resonance.

In this paper we have assumed that the orbital motion of gas within
the disk deviates from perfect circles only to the minimal extent
required by tidal forces.  For most systems this is a reasonable
assumption, but there are particularly cases where it is likely
to fail.  For example, SU Uma stars exhibit superoutbursts which
are probably due to the expansion of the outer edge of the
disk through the $3:1$ orbital resonance (\cite{w88}).  Lubow (1991a, 1991b)
has shown that at this radius the disk is vulnerable to
an elliptical instability.  Under these conditions the tidal parametric
instability discussed here can exist as a secondary instability,
implying a dramatic rise in the generation of internal waves during
a superoutburst.

\acknowledgements
This work has been supported in part by NASA grant NAG5-2773.
In addition, the author would like to thank the Harvard-Smithsonian
Center for Astrophysics for their hospitality while this paper
was being written and Jeremy Goodman and Dongsu Ryu for several
helpful conversations.

\clearpage
\begin{figure}
\plotone{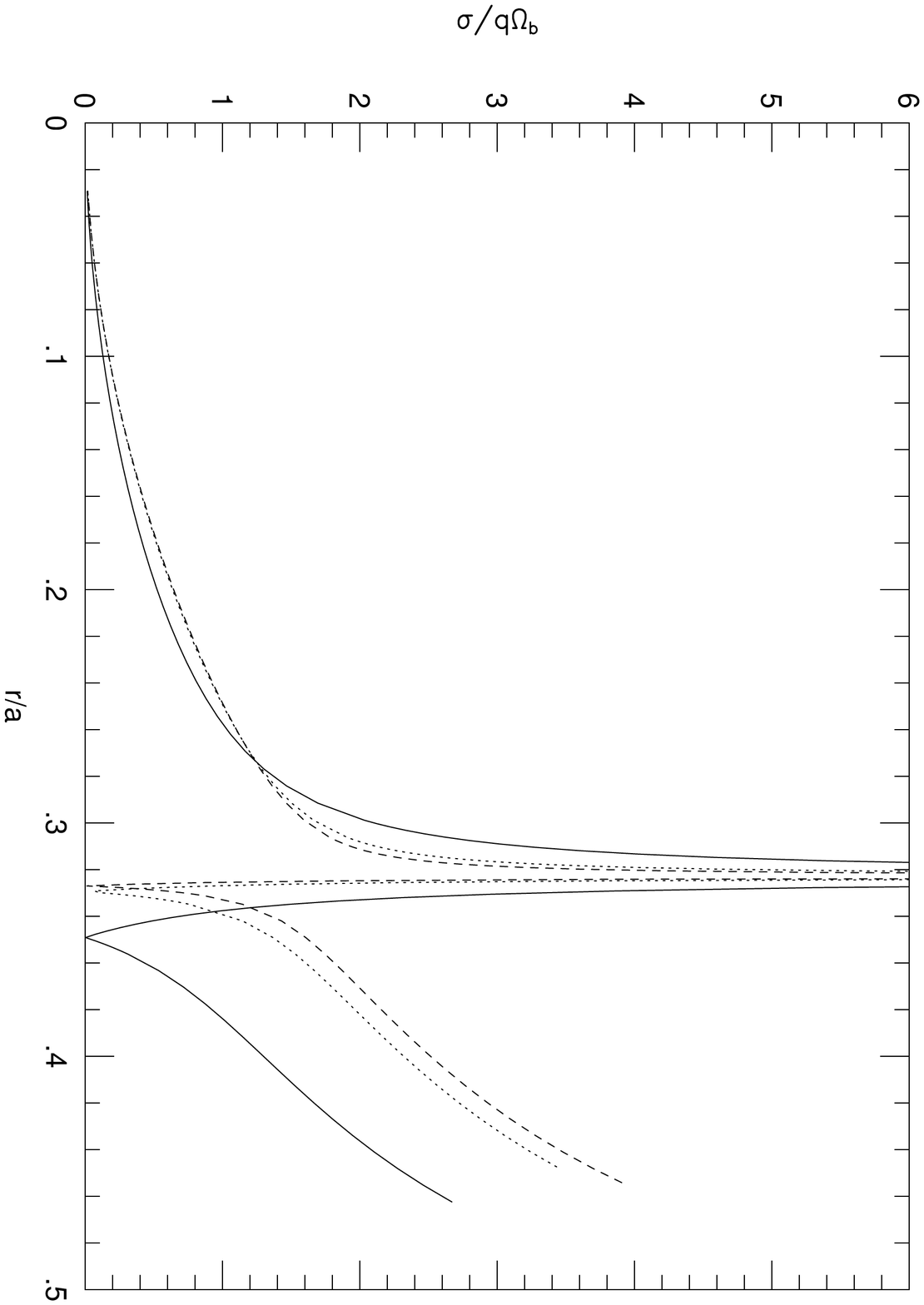}
\caption{Growth rates for the parametric instability in units of
$q\Omega_b$ as a function of $r/a$ for an adiabatic accretion disk
with $\gamma=5/3$. The solid line is for the minimal vertical
wavenumber.  The dotted and dashed lines are for the second and third
wavenumbers.}
\end{figure}
\begin{figure}
\plotone{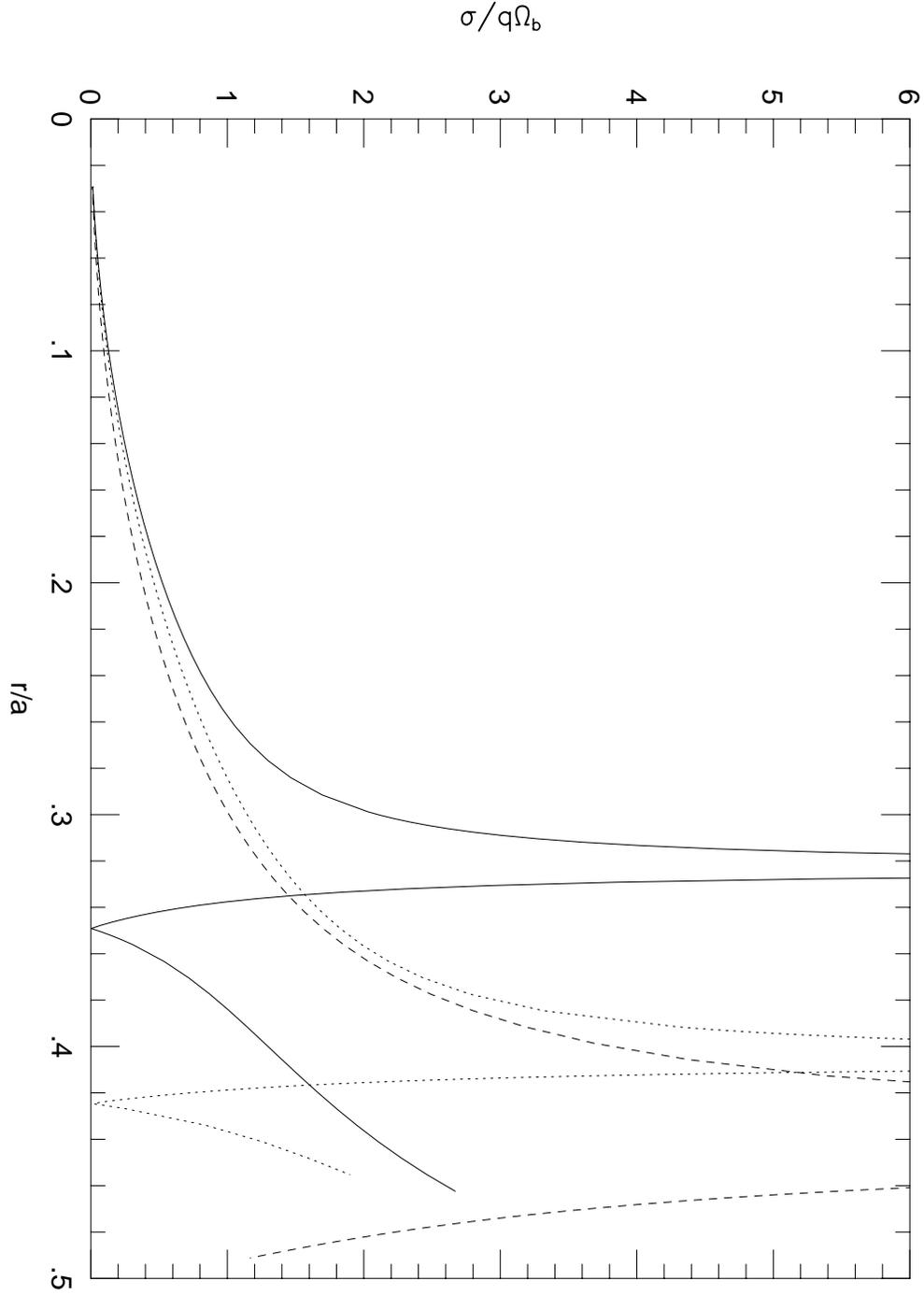}
\caption{Growth rates for the minimal vertical wavenumber mode in an
adiabatic disk for $\gamma=5/3$ (solid line), $\gamma=1.2$ (dotted line),
and $\gamma=1.0$ (dashed line).}
\end{figure}
\begin{figure}
\plotone{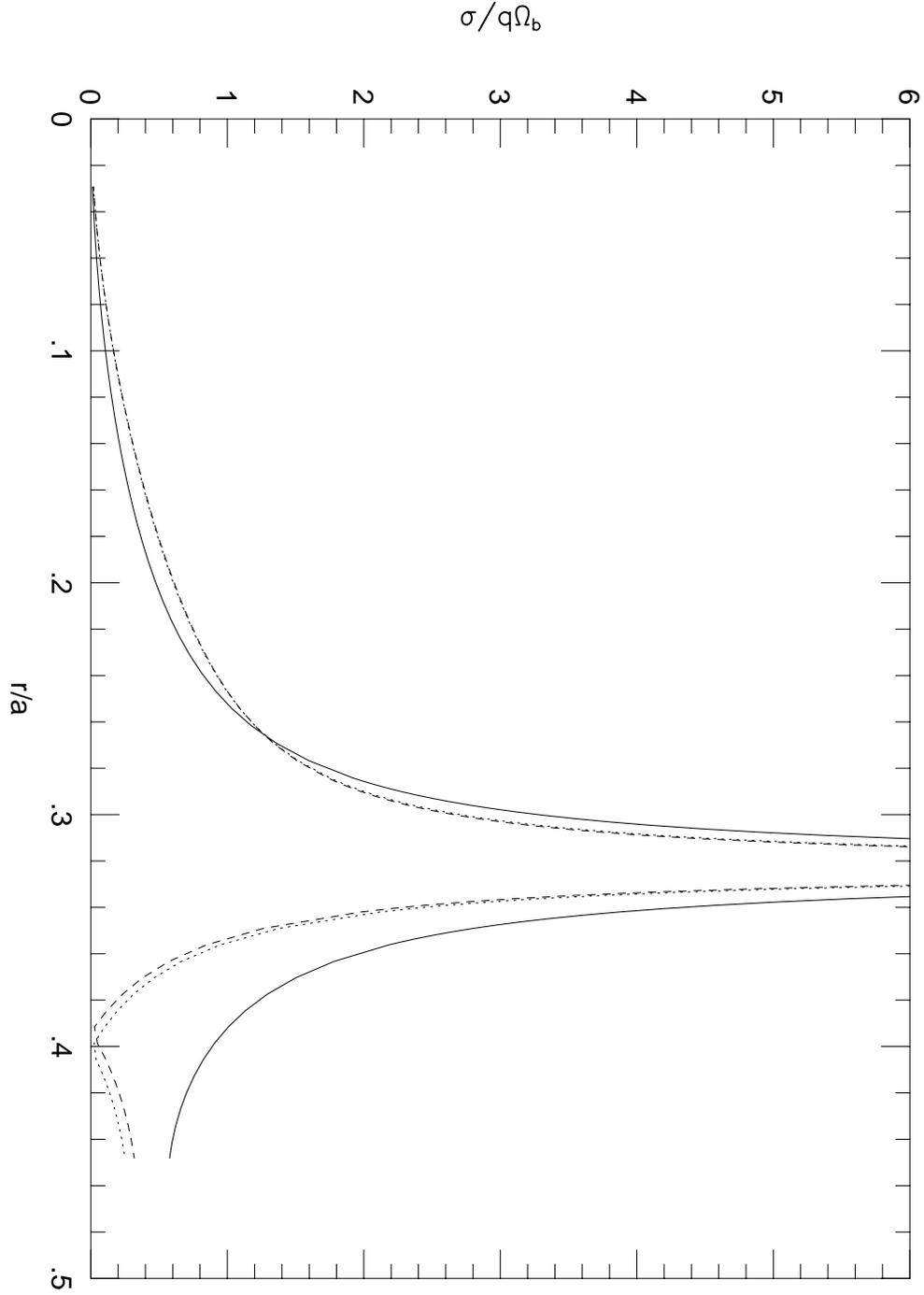}
\caption{Growth rates for an accretion disk with an isothermal vertical
structure and $\gamma=5/3$.  The solid line is for the minimal vertical
wavenumber.  The dotted and dashed lines are for the second and third
wavenumbers.}
\end{figure}
\begin{figure}
\plotone{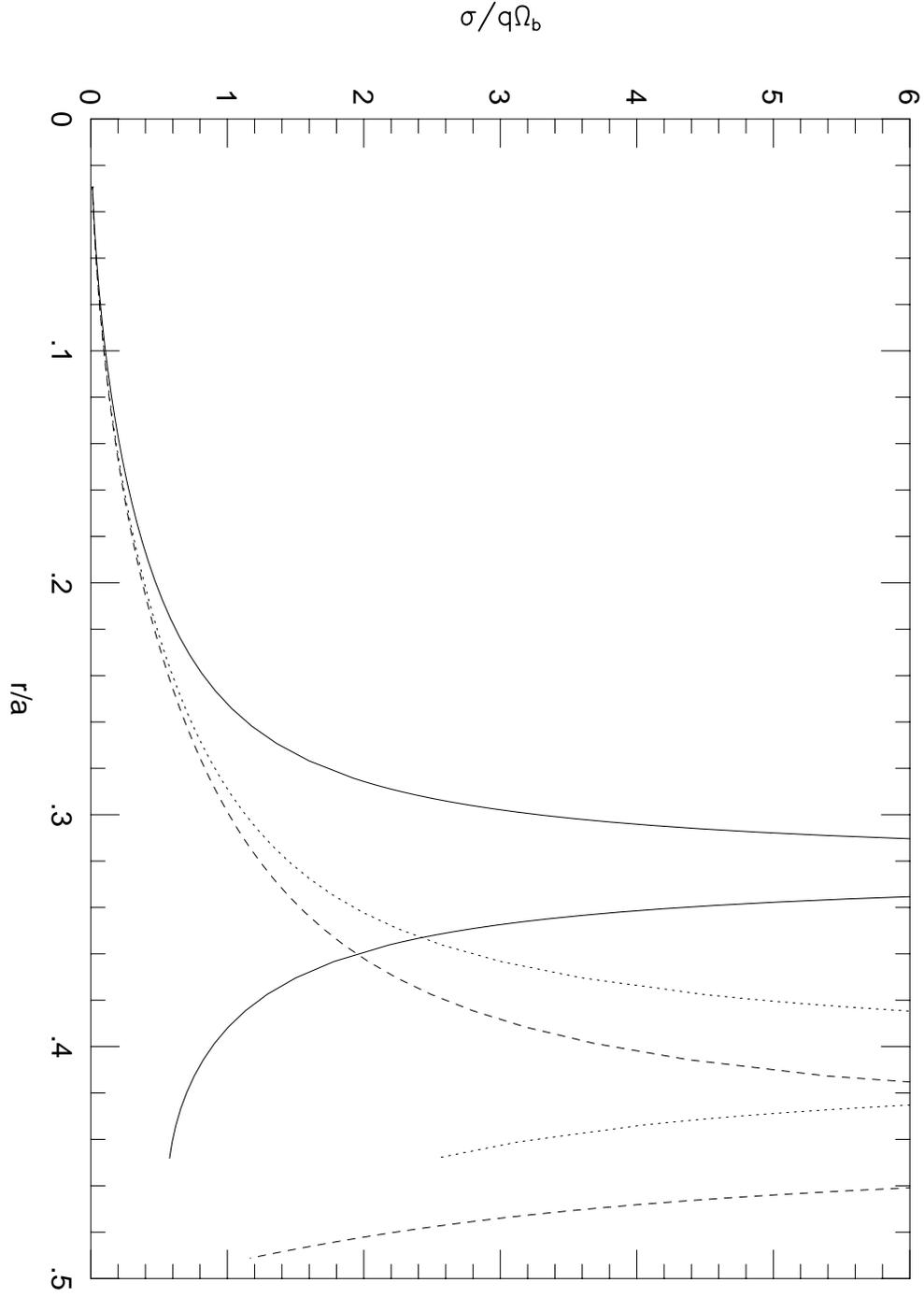}
\caption{Growth rates for the minimal vertical wavenumber mode in an
isothermal disk for $\gamma=5/3$ (solid line), $\gamma=1.2$ (dotted line),
and $\gamma=1.0$ (dashed line).}
\end{figure}
\begin{figure}
\plotone{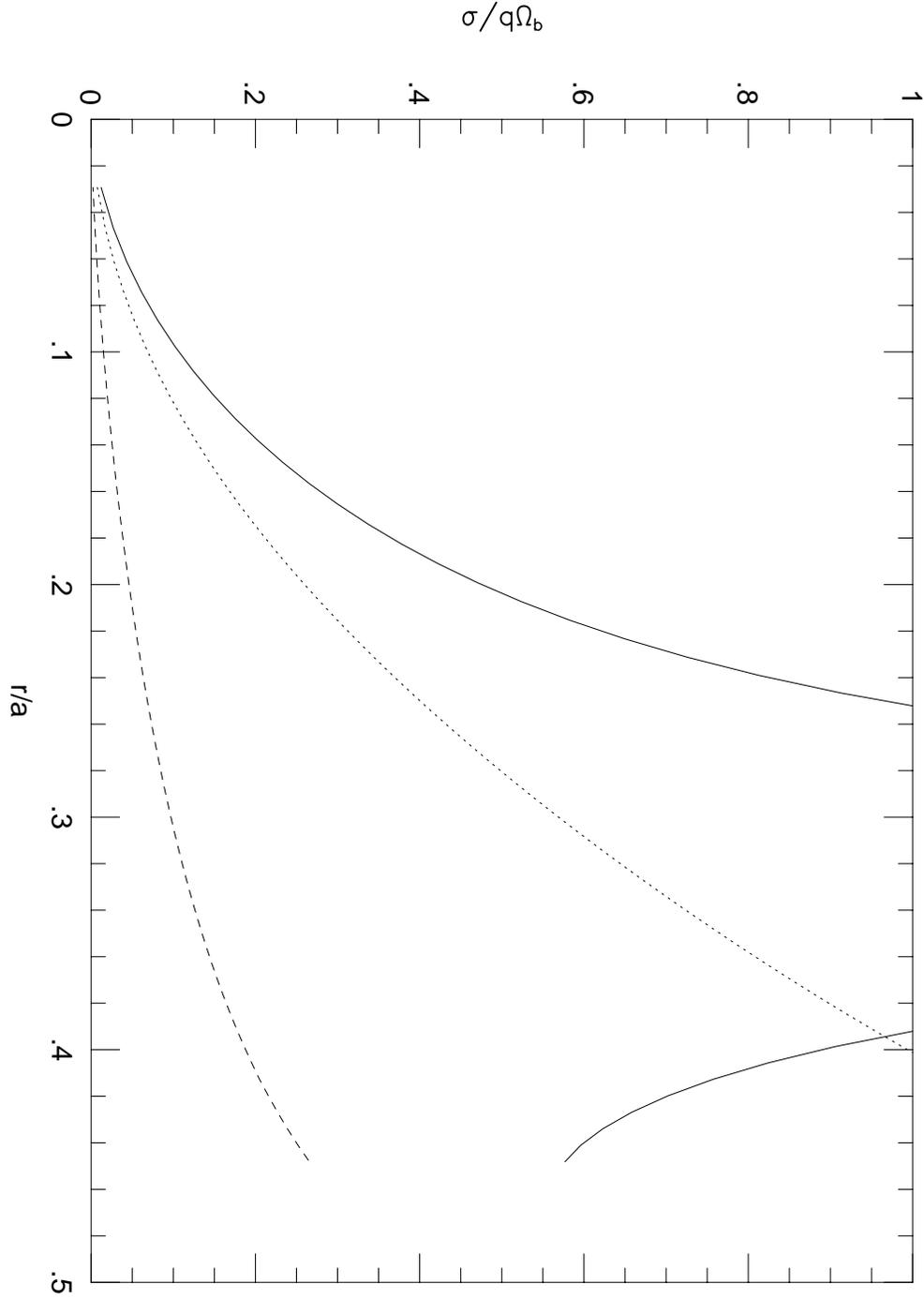}
\vskip -1cm
\caption{Growth rates for the minimal vertical wavenumber mode in
an isothermal disk with $\gamma=5/3$.  The solid line is full result.
The dotted line shows the effect of suppressing vertical motion and
tidal perturbations to density and pressure.  The dashed line results
from also assuming $\vec\xi\propto r$.}
\end{figure}
\end{document}